\documentclass[onecolumn,showpacs,amsmath,amssymb,pra,superscriptaddress,floatfix]{revtex4}
\usepackage{graphics}
\usepackage{graphicx}
\begin{document}
\title{
Matter-Wave Solitons in the Presence of Collisional Inhomogeneities: Perturbation theory and the impact of derivative terms}
\date{\today}

\author{S. Middelkamp}
\email[]{stephan.middelkamp@pci.uni-heidelberg.de}
\affiliation{%
Theoretische Chemie, Physikalisch-Chemisches Institut,
Universit\"at Heidelberg,
INF 229, 69120 Heidelberg, Germany}%
\author{P.G.\ Kevrekidis }
\email[]{kevrekid@math.umass.edu}
\affiliation{Department of Mathematics and Statistics, University of Massachusetts,
Amherst MA 01003-4515, USA}
\author{D.J.\ Frantzeskakis}
\email[]{dfrantz@phys.uoa.gr}
\affiliation{Department of Physics, University of Athens, Panepistimiopolis, Zografos, Athens 157 84, Greece}
\author{P. Schmelcher}
\email[]{Peter.Schmelcher@pci.uni-heidelberg.de}
\affiliation{%
Theoretische Chemie, Physikalisch-Chemisches Institut,
Universit\"at Heidelberg,
INF 229, 69120 Heidelberg, Germany}%
\affiliation{%
Physikalisches Institut, Universit\"at Heidelberg, Philosophenweg 12, 69120 Heidelberg, Germany}%

\date{\today}

\begin{abstract}\label{txt:abstract}
We study the dynamics of bright and dark matter-wave solitons 
in the presence of a spatially varying nonlinearity. 
When the spatial variation does not involve zero crossings, a 
transformation is used to bring the problem to a standard 
nonlinear Schr{\"o}dinger form, but with two additional terms:
an effective potential one and a non-potential term. We illustrate
how to apply perturbation theory of dark and bright solitons to
the transformed equations. We develop
the general case, but primarily focus on the non-standard special
case whereby the potential term vanishes, for an inverse 
square spatial dependence of the nonlinearity. In both 
cases of repulsive and attractive interactions, 
appropriate versions of the soliton perturbation theory 
are shown to accurately describe the soliton dynamics.
\end{abstract}
\maketitle

\section{Introduction}

%
The experimental creation of atomic Bose-Einstein condensates (BECs) 
has been one of the most fundamental developments in quantum and atomic physics
over the past two decades. The impressive progress in this field due 
to intense experimental and theoretical studies has been 
already summarized in various books \cite{book1,book2,Kevrekidis} and reviews \cite{ricardo}. 
This progress has been, to a considerable extent, fueled by the 
fact that, in a mean-field picture, BECs can be described by a macroscopic 
wavefunction obeying 
%
the Gross-Pitaevskii (GP) equation, which is an equation of the nonlinear 
Schr{\"o}dinger (NLS) type. In such a mean-field description, 
the effective nonlinearity (which is introduced by interatomic interactions) 
%
%
allows for studies of macroscopic nonlinear matter waves; in this respect it is important to note 
that bright matter-wave solitons in attractive BECs \cite{expb1,expb2,expb3}, 
as well as dark \cite{dark1,dark2,dark3,dark4,dark5,dark6,dark7} 
and gap \cite{gap}
matter wave solitons in repulsive BECs, have been observed in a series of experiments  
(see also the recent review \cite{ricardo}).

One of the remarkable possibilities arising in the physics of BECs is that 
%
%
the interatomic interactions (and, hence, the effective nonlinearity) 
can be manipulated by means of different types of temporally- or spatially-varying external potentials.
More specifically, the s-wave scattering length (which is proportional
to the nonlinearity coefficient in the GP equation) can be
experimentally adjusted using either magnetic \cite{Koehler,feshbachNa} or 
optical Feshbach  resonances \cite{ofr} in a very broad range.
The availability of these tools has led to a number of consecutive theoretical and experimental studies. For instance, the formation of bright 
matter-wave solitons and soliton trains of 
$^{7}$Li \cite{expb1,expb2} and $^{85}$Rb \cite{expb3} atoms
used a tuning of the interatomic interactions from repulsive to attractive. 
Also, this type of manipulations was instrumental 
in achieving the formation of molecular condensates \cite{molecule}, 
and the probing of the BEC-BCS crossover \cite{becbcs}. 
A parallel track of theoretical studies has explored the use 
of a time-dependent modulation of the nonlinearity coefficient to
stabilize attractive higher-dimensional BECs against collapse \cite{FRM1}, or
to create robust matter-wave breathers in lower-dimensional BECs \cite{FRM2}. 
More recently, the use of spatial variations of the nonlinearity 
to create so-called ``collisionally inhomogeneous'' environments has been proposed. 
In that regard, 
major developments included 
adiabatic compression of matter-waves \cite{our1,fka}, 
Bloch oscillations of matter-wave solitons \cite{our1},  
atomic soliton emission and atom lasers \cite{vpg12}, 
enhancement of transmittivity of matter-waves through barriers 
\cite{our2,fka2}, dynamical trapping of matter-wave solitons \cite{our2}, 
stable condensates exhibiting both attractive and repulsive interatomic 
interactions \cite{chin} as well as the 
delocalization transition of matter waves~\cite{LocDeloc}.
Among the types of spatial variations of the nonlinearity that
have been proposed, one can trace linear ones \cite{our1,our2}, as well as 
parabolic \cite{yiota}, random \cite{vpg14}, 
periodic \cite{vpg16,LocDeloc,BludKon,augusto}, and 
localized (step-like) \cite{vpg12,vpg17,vpg_new} 
ones. On the mathematical side, a number of detailed
studies \cite{key-2,key-4,vprl} have appeared, addressing aspects
such as the effect of a ``nonlinear lattice potential" (i.e., a spatially 
periodic nonlinearity) on the stability
of 
matter-wave solitons, and the interplay between drift and diffraction/blow-up 
instabilities. 
More recently, the interplay of nonlinear and linear potentials has been 
examined in both continuum \cite{ckrtj} and discrete \cite{blud_pre} settings
(see also the recent work \cite{hiza} and references therein).

Our aim in this work is to 
study the dynamics of matter-wave solitons 
in the presence of a spatially-dependent nonlinearity. We consider both
dark solitons in 
repulsive BECs, 
as well as bright solitons in 
attractive BECs. 
In the case where the sign of the nonlinearity coefficient (hereafter
referred to as $g(x)$) does not change, we first show that a change of
variables can convert the spatially variable nonlinearity problem into
a ``regular'' one where the nonlinearity has a 
constant prefactor. This transformation results in the emergence of 
two additional perturbation terms:
one of them can be considered as an effective potential term 
(i.e., a spatially-dependent function multiplying the 
macroscopic wavefunction $u$), while
the other one can not (it consists of a spatially-dependent function
multiplying the {\it derivative} of the 
wavefunction $\partial_x u$, for an elongated BEC along the $x$-direction).
We use this transformation as a starting point in order
to develop perturbation theory for the soliton dynamics 
in the presence of $g(x)$ for the case of arbitrary $g(x)$.
However our focus is on the case where $g(x)$ is such that
the potential term completely vanishes. The reason for this selection
is that it appears to be the less physically intuitive case (due to
the derivative nature of the corresponding perturbation). Moreover, 
the effect of a perturbation induced by a ``standard'' potential term has been 
studied fairly extensively in the BEC context (see, e.g., \cite{book1,book2,Kevrekidis,ricardo}), 
also in the particular case of collisionally inhomogeneous BECs (see, e.g., 
\cite{our1,fka,our2}). 

Our investigation is structured as follows. In the next section, we give the general
setting and analyze the relevant transformation. In section III,
we focus on dark solitons, first providing the general theory, and
then applying it to the particular case of interest. In section IV,
we follow a similar path for the case of bright solitons. Finally,
in section V, we summarize our findings and present our conclusions,
as well as some interesting directions for future study.

\section{Perturbed Gross-Pitaevskii Equation and the derivative-only case}

\begin{figure}[htbp]
\includegraphics[angle=270,width=7cm]{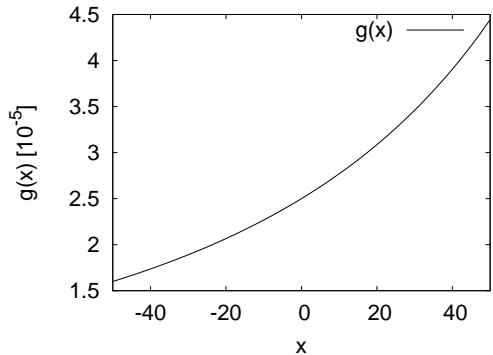}  
  \caption{Spatial dependence of the interaction parameter $g(x)$ for $C=1$ and $D=-200$ for a repulsive condensate.}  
\label{fig: g(x)}
\end{figure}

In this work we will restrict ourselves to an effective one-dimensional (1D) 
description accounting only for the longitudinal dynamics of the condensate. 
In the transverse directions the atoms should be tightly confined which can be 
realized by an isotropic harmonic potential with a trap frequency $\omega_\perp$ 
(associated with the harmonic oscillator length $a_\perp$). For $a_\perp$ 
small enough, one can regard the transversal dynamics as frozen
(see chap. 1 in \cite{Kevrekidis} and also \cite{perez-garcia, Jackson}),  
i.e., only the corresponding ground state is occupied. 
The longitudinal motion takes place in the $x$-direction and 
should not be confined. Then, the corresponding mean-field equation is given by 
\begin{equation}
 i \hbar\partial_t \Psi=-\frac{\hbar^2}{2m}\partial^2_x\Psi+g(x)|\Psi|^2\Psi
\label{gp}
\end{equation}
where $\Psi(x,t)$ is the macroscopic wave function, 
$m$ the atomic mass and $g(x)=g^{(3D)}(x)/2\pi a_{\perp}^2$ the effective 1D interaction coefficient. 
The parameter $g^{(3D)}(x)=4\pi\hbar^2 a/m$ characterizes the two-particle interaction in 3D 
with the s-wave scattering length $a$. The 
latter is positive (negative) for repulsive (attractive) condensates consisting of, e.g., $^{23}$Na ($^7$Li) atoms. 
The value of the scattering length can be tuned, as 
mentioned above, e.g., by use of magnetic Feshbach resonances \cite{Marte02}. In the vicinity of a 
magnetic Feshbach resonance the value of the scattering length depends on the value of an applied 
magnetic field. Thus, one can achieve a spatially dependent scattering length by applying an 
inhomogeneous magnetic field yielding a collisionally inhomogeneous BEC. We now 
use 
suitable straightforward rescalings (see, e.g., \cite{our1}) and dimensionless units to 
express Eq. (\ref{gp}) in the following form:
\begin{equation}
 i \partial_t \Psi=-\frac{1}{2}\partial^2_x\Psi+s|g(x)||\Psi|^2\Psi, 
\label{eq: GP normalized}
\end{equation}
where the coefficient $s={\rm sgn}(g)=\pm 1$ for attractive and repulsive condensates, respectively. 
Applying the transformation $\Psi=\frac{u}{\sqrt{g}}$ allows us to rewrite eq. (\ref{eq: GP normalized}) in the following way:
\begin{equation}
i\partial_t u= -\frac{1}{2}\partial^2_x u+s|u|^2 u + \tilde V_{eff}(x) u -\sqrt{g}\partial_x\frac{1}{\sqrt{g}}\partial_x u, 
\label{eq: GP transformed}
\end{equation}
with the effective potential term $\tilde V_{eff}(x)=-\frac{1}{2}\sqrt{g}\partial^2_x\frac{1}{\sqrt{g}}$. 
Equation (\ref{eq: GP transformed}) can be written as the usual NLS equation 
(with a defocusing or focusing nonlinearity for $s=\pm 1$, respectively) 
with an external spatially-dependent perturbation $P[u(x,t);x]$, namely:
\begin{equation}
i\partial_t u +\frac{1}{2}\partial^2_x u-s|u|^2 u=P[u(x,t);x].
\label{eq: GP perturbed}
\end{equation}
The perturbation can be expressed as $P[u(x,t);x]=P_{L}[u(x,t);x]+P_{NP}[u(x,t);x]$, i.e., 
it consists of a linear effective potential contribution 
$P_{L}[u(x,t);x]=\tilde V_{eff}(x)u(x,t)$, as well as of a non-potential 
perturbation of the form $P_{NP}[u(x,t);x]=-\sqrt{g}\partial_x\frac{1}{\sqrt{g}}\partial_x u$. 
In this work we are mainly interested in the effects of the 
less standard, non-potential type of perturbation. 
Therefore, we assume the collisionally inhomogeneous interaction to be of the form
\begin{equation}
g(x)=\frac{1}{(D+Cx)^2}, 
\label{g(x)}
\end{equation}
with arbitrary constants $C$ and $D$. For such a particular selection
of $g(x)$, $\tilde V_{eff}(x)$ vanishes leading to the perturbation 
\begin{equation}
P[u(x,t);x] = -\frac{C}{D+Cx}  \partial_x u,
\label{eq: perturbation} 
\end{equation}
consisting only of the non-potential contribution in the right hand side. 
Thus we can investigate the pure effects of the latter, non-standard 
contribution (the effects of a standard linear potential have been
studied fairly extensively; see e.g. \cite{book1,book2,Kevrekidis,ricardo}). 
We choose $C=1$ and $D=-200$; thereby the 
singularity in the perturbation occurs at $x_0^{sing}=200$ (which will be 
outside the region of interest in our domain). For this choice, the spatial dependence 
of the coefficient $g(x)$ is shown in Fig. \ref{fig: g(x)}.

\section{Dark Matter-Wave Solitons}
\subsection{Full perturbative approach}
Let us first consider the case of dark matter-wave solitons for $s=1$. 
In order to treat effects of the perturbation (\ref{eq: perturbation}) on a 
dark soliton analytically we employ the adiabatic perturbation theory 
assuming that the functional form of the soliton remains unchanged by the 
perturbation (an assumption that in our setting will be justified
a posteriori). We first use the transformation $u \rightarrow u\exp(-it)$ to put 
Eq. (\ref{eq: GP perturbed}) in the form $i\partial_t u +\frac{1}{2}\partial^2_x u-(|u|^2 -1) u=P[u(x,t);x]$ 
and use as an ansatz for the soliton, the following expression,
%
\begin{equation}
u=B\tanh(B(x-x_0))+i A, 
\label{eq: dark soliton unperturbed}
\end{equation}
which is the exact dark soliton solution of the above mentioned unperturbed NLS equation.
According to the above discussion, the soliton depth $A$, velocity $B$ (with $A^2+B^2 =1$), and center $x_0$ 
are assumed to be unknown functions of 
time. 
The Lagrangian density of an unperturbed dark soliton is 
given by \cite{Theocharis05}:
\begin{equation}
\mathcal{L}(u)=\frac{i}{2}(u^{\star}\partial_t u-u\partial_t u^{\star})
(1-\frac{1}{|u|^2})-\frac{1}{2}|\partial_x u|^2-\frac{1}{2}(|u|^2-1)^2,  
\label{eq: Lagrangian density}
\end{equation}
while the averaged Lagrangian, $L=\int dx \mathcal{L}(u)$, can be calculated 
by substituting (\ref{eq: dark soliton unperturbed}) in  Eq. (\ref{eq: Lagrangian density}) 
yielding:
\begin{equation}
 L(A,x_{0})=2\partial_t x_0 \bigl(-AB+\tan^{-1}(\frac{B}{A})\bigr)-\frac{4}{3}B^3.
\end{equation}
In \cite{Kivshar95} it was shown that, within the framework of the
adiabatic perturbation theory for small perturbations, 
the parameters of the soliton $\alpha_j=\lbrace x_0,A\rbrace$ obey the 
the Euler-Lagrange equations
\begin{equation}
 \partial_{\alpha_j}L-\frac{d}{dt}\partial_{\alpha^{\prime}_j}L=2Re\lbrace\int dx P^{\star}(u)\partial_{\alpha_j}u\rbrace
\end{equation}
with $\alpha_j^{\prime}=\partial_t \alpha_j$. This leads to a system of ordinary differential equations (ODE) for $A$ and $x_0$:
\begin{eqnarray}
\partial_t A&=&\frac{1}{2}B^3\int dx \sqrt{g}\partial_x \frac{1}{\sqrt{g}} {\rm sech}^4(B(x-x_0))\nonumber\\
&&+\frac{1}{4}B^2\int dx \sqrt{g}\partial^2_x\frac{1}{\sqrt{g}}\tanh(B(x-x_0)) {\rm sech}^2(B(x-x_0)), 
\label{eq: A general}\\
\partial_{t}x_0&=&A-\frac{1}{2}A\int dx \sqrt{g}\partial_x \frac{1}{\sqrt{g}}{\rm sech}^2(B(x-x_0))\Bigl(\tanh(B(x-x_0))\nonumber\\
&&+B(x-x_0) {\rm sech}^2(B(x-x_0))\Bigr)\nonumber\\
&&-\frac{1}{4}\int dx \sqrt{g}\partial^2_x\frac{1}{\sqrt{g}}\Bigl((\frac{1}{B} \tanh^2(B(x-x_0))-1)\nonumber\\
&&+(x-x_0)\tanh(B(x-x_0)) {\rm sech}^2(B(x-x_0))\Bigr).
\label{eq: x_0 general}
\end{eqnarray}
For an interaction obeying Eq. (\ref{g(x)}), 
the terms arising from the linear potential vanish, leading to the system:
\begin{eqnarray}
\partial_t A&=&\frac{1}{2}B^3\int dx \frac{C}{D+Cx} {\rm sech}^4(B(x-x_0)),
\label{eq: A}\\
\partial_{t}x_0&=&A-\frac{1}{2}A\int dx \frac{C}{D+Cx} {\rm sech}^2(B(x-x_0))\Bigl(\tanh(B(x-x_0))\nonumber\\
&&+B(x-x_0) {\rm sech}^2(B(x-x_0))\Bigr). \label{eq: x_0}
\end{eqnarray}

\subsection{Approximations}

We can simplify the general framework of Eqs. (\ref{eq: A general},\ref{eq: x_0 general}) by performing a 
Taylor expansion of the interaction around $x=x_0$, leading in first order to 
\begin{eqnarray}
 \partial_t A&\approx&\frac{2}{3}(1-A^2)\sqrt{g(x_0)}\partial_x \frac{1}{\sqrt{g(x)}}\Bigm|_{x=x_0}\label{eq: dark taylor A general}\\
 \partial_t x_0&\approx&A+\frac{1}{4}\frac{A}{B^2}\sqrt{g(x_0)}\partial_x^2\frac{1}{g(x)}\Bigm|_{x=x_0}. \label{eq: dark taylor x_0 general}
\end{eqnarray}
The Taylor expansion around $x_0$ can be justified in most settings 
due to the exponential localization of the soliton around its center. 
Dropping higher-order terms essentially implies that the interaction does not 
change on the scale of the width of the soliton (if the latter assumption
is invalid, then we can not resort to this approximation). 
In the special case of the interaction (\ref{g(x)}) the following evolution 
equations are obtained:
\begin{eqnarray}
 \partial_t A&\approx&\frac{2}{3}(1-A^2)\frac{C}{D+Cx_0}\label{eq: dark taylor A}\\
 \partial_t x_0&\approx&A
\label{eq: dark taylor x_0}
\end{eqnarray}
By combination of Eqs. (\ref{eq: dark taylor A}) and (\ref{eq: dark taylor x_0}), we obtain a single second-order 
ordinary differential equation (ODE) 
\begin{equation}
\partial^2_t x_0 = \frac{2}{3}\frac{C}{D+Cx_0}(1-(\partial_t x_0)^2) \label{eq: dark EOM}
\end{equation}
for the center of the soliton. If the velocity of the soliton is small,  
$\partial_t x_0 \ll 1$, one can neglect the second term in 
the right hand side of Eq. (\ref{eq: dark EOM}) leading to: 
\begin{equation}
 \partial^2_t x_0 = \frac{2}{3}\frac{C}{D+Cx_0}. \label{eq: dark EOM appr}
\end{equation}
We will discuss the validity of this approximation in our numerical
results below. Equation (\ref{eq: dark EOM appr}) is the equation of motion (EOM) 
\begin{equation}
\partial^2_t x_0 =-\partial_{x_0} V^{eff}(x_0),
\end{equation}
of a particle 
in the presence of the effective potential
\begin{eqnarray}
 V^{eff}(x_0)&=&-\frac{2}{3}\ln(|Cx_0+D|). 
\label{eq: dark effective pot}
\end{eqnarray}
Therefore, we will denote Eq. (\ref{eq: dark EOM}) as EOM and Eq. 
(\ref{eq: dark EOM appr}) as EOM$_a$ in the next section.
As an interesting aside, we should note that 
even in the presence of the kinetic term (i.e., if $(\partial_t x_0)^2$ is 
not neglected), one can rewrite Eq. (\ref{eq: dark EOM}) as a Hamiltonian 
system (see Ref. \cite{Smereka87}) using a generalized momentum 
$P=g(x_0)\partial_t x_0$. With this momentum one finds for a system 
\begin{equation}
 \partial^2_t x_0 = f(x_0)(1-(\partial_t x_0)^2), 
\end{equation}
the equations of motion
\begin{eqnarray}
  \partial_t x_0 &=&\frac{P}{g(x_0)}, \\
  \partial_t P &=& g(x_0)f(x_0)+P^2 \bigl(\frac{\partial_{x_0}g(x_0)}{(g(x_0))^2}-\frac{f(x_0)}{g(x_0)}\bigr), 
\end{eqnarray}
which correspond to the Hamiltonian
\begin{equation}
 H(x_0,P)=\frac{1}{2}\frac{P^2}{g(x_0)}+F(x_0),
\end{equation}
with 
\begin{eqnarray}
 g(x_0)&=&A \exp(2\int^{x_0}dx_{0}^{\prime}f(x_0^{\prime})), 
\\
 F(x_0)&=&-\frac{1}{2}g(x_0).
\end{eqnarray}
In the particular case of $f(x_0)=\frac{2}{3}\frac{C}{D+Cx_0}$, 
one obtains for the momentum
\begin{equation}
 P=A(D+Cx_0)^{\frac{4}{3}}\partial_t x_0,
\end{equation}
and for the Hamiltonian
\begin{equation}
 H(x_0,P)=\frac{P^2}{2A}(D+Cx_0)^{-\frac{4}{3}}-\frac{1}{2}A(D+Cx_0)^{\frac{4}{3}}.
\end{equation}

\subsection{Numerical Results}

In this section we present and compare the numerical results obtained by 
solving the full partial differential equation (PDE) of the GP type 
(\ref{eq: GP perturbed}), as well as the ODEs 
(\ref{eq: A},\ref{eq: x_0}), the EOM (\ref{eq: dark EOM}) and the simplified 
EOM$_{a}$ (\ref{eq: dark EOM appr}). We have confirmed that 
throughout our simulations the soliton is localized in a region with a 
well defined perturbation avoiding the singularity. Since the soliton is 
exponentially localized, the spatial integrations in Eqs. (\ref{eq: A},\ref{eq: x_0}) 
can be restricted to a region around the center of the soliton with a finite 
perturbation and a well defined integrand. The time evolution is performed by 
the Adams-Bashforth-Moulton predictor-corrector method.

\begin{figure}[htbp]
\includegraphics[angle=270,width=7cm]{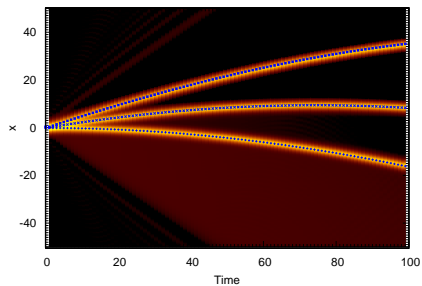}  
  \caption{Time evolution of the density of dark solitons with $x_0(0)=0$ 
and $A_{init}=0$, $0.25$, $0.5$ (from bottom to top). The dotted line is the 
CM parameter ($x_0$) obtained by solving the ODEs 
(\ref{eq: A})-(\ref{eq: x_0}).}  
   \label{fig: dark density}
\end{figure}

Figure \ref{fig: dark density} shows the time evolution of the density profile 
of a dark soliton with $x_0(0)=0$ for different initial velocities obtained 
by solving Eq. (\ref{eq: GP perturbed}). Black represents the highest density,
while white corresponds to the lowest density. 
The dotted lines are the results for $x_0$ (the center of the soliton) 
as obtained by solving the ODEs (\ref{eq: A},\ref{eq: x_0}). The results 
agree very well, showing that the adiabatic perturbation theory describes the 
motion of the center of the soliton accurately. For $A_{init}=0$ (lowest curve)
the soliton gets accelerated to the negative half-plane and moves immediately 
into this direction. For $A_{init}=0.25$ and $A_{init}=0.5$ (middle and top 
curve, respectively), the soliton also gets accelerated into the direction of 
the negative half-plane but starts moving to the positive one due to its
initial velocity until it reaches a turning point of zero velocity and changes 
direction. By investigation of Fig. \ref{fig: g(x)} one observes that in the 
considered region the value of the interaction parameter decreases for 
decreasing $x$. 
Thus, the soliton gets accelerated into the direction with a 
smaller interaction parameter. Due to the fact that the interaction is repulsive, the 
interaction energy decreases with decreasing interaction parameter. 
So the soliton tends to move into the region with less interaction energy. This 
happens despite the fact that the interaction parameter does not enter 
explicitly in the perturbation as a potential but rather through 
the product of its first derivative and the first derivative of the scaled wavefunction.

\begin{figure}[htbp]
  \begin{minipage}[c]{7 cm}
\includegraphics[angle=270,width=7cm]{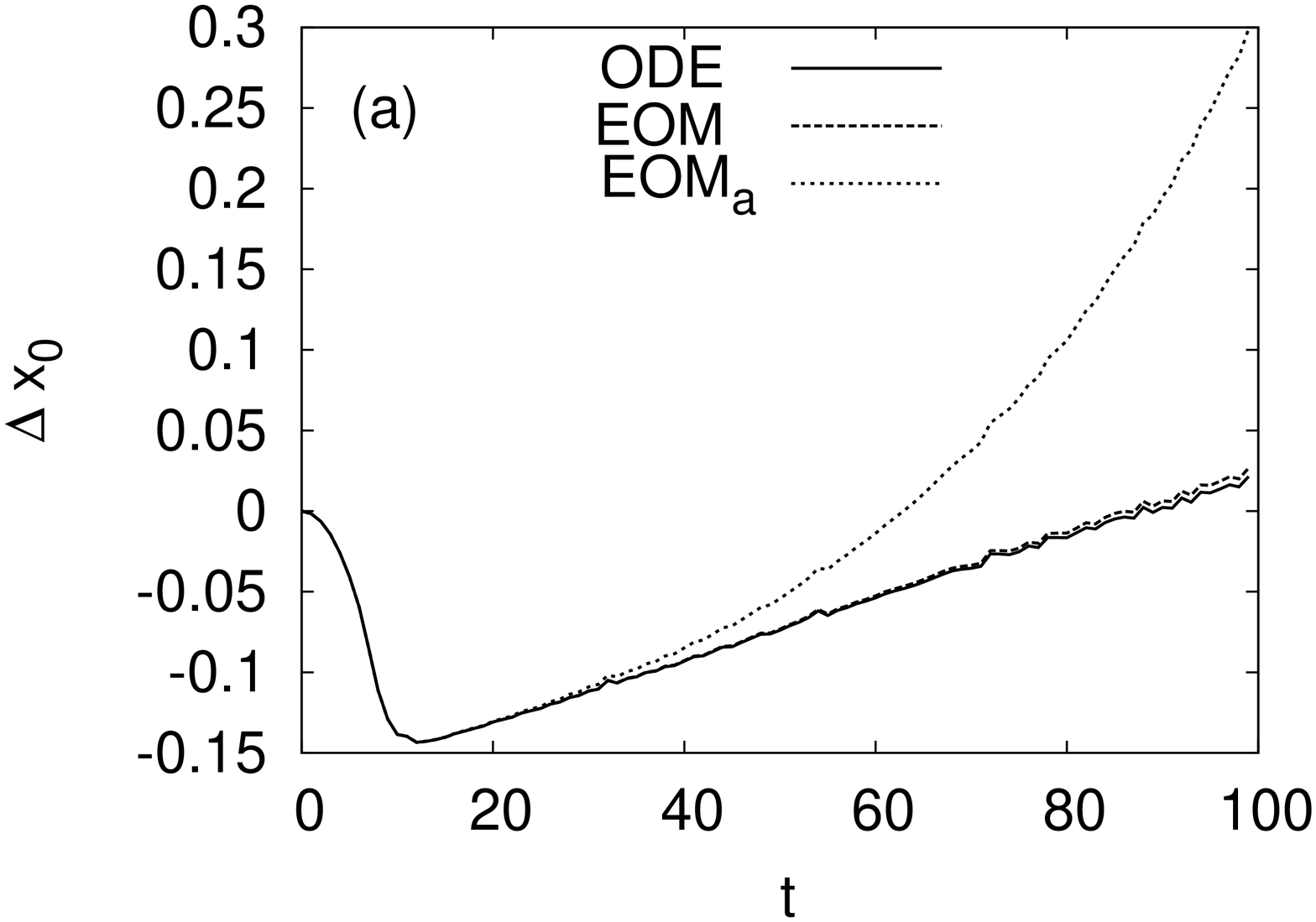}  
  \end{minipage}
  \begin{minipage}[c]{7 cm}
\includegraphics[angle=270,width=7cm]{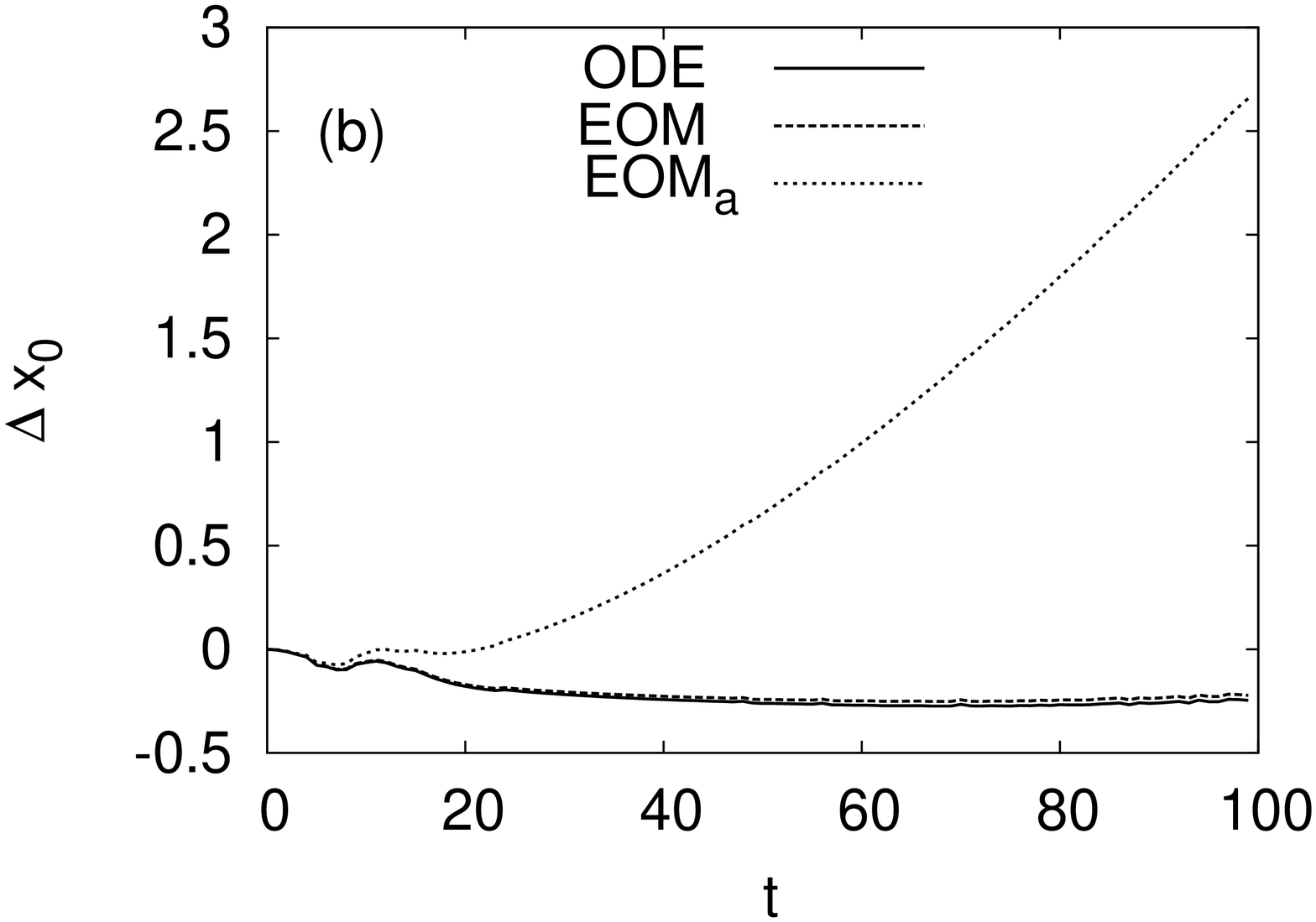}  
  \end{minipage}
  \caption{Difference of the soliton center $\Delta x_0$ calculated by solving the PDE with the result of the ODEs, EOM and 
EOM$_{a}$ for  $A_{init}=0.0$ (a) and $A_{init}=0.5$ (b).}
\label{fig: dark deltax}
\end{figure}

Figure \ref{fig: dark deltax} shows the time evolution for $x_0(0)=0$, 
$A_{init}=0$(a) and $A_{init}=0.5$(b) of the differences between the results 
for the center of the soliton obtained by solving the PDE, and the ODE, EOM 
and EOM$_{a}$, respectively. We calculated the center of mass 
of the PDE solution by performing the integration 
$x_0=\int x(b-|u|^2)dx / \int (b-|u|^2)dx $, with 
$b=|u(x_b)|^2$ being the background density evaluated far away from the center 
of the soliton. The differences for the ODEs and the EOM are almost equal and 
small for both initial velocities. So the adiabatic perturbation theory works 
fine for describing the center of the soliton. The results obtained by solving 
EOM$_{a}$ coincide for a small time period with the result of the PDE. For 
longer times, they deviate from these results. For a larger initial velocity 
the deviation is even larger. The reason for this is that we neglected the 
impact of the velocity in EOM$_a$ and, thus, 
the approximation gets worse for larger velocities. However, the qualitative 
behavior is described correctly even within  this approximation. 
Hence, one can understand 
the behavior of the soliton as a particle moving in the effective potential (\ref{eq: dark effective pot}) and 
thus can explain the acceleration observed in Fig. \ref{fig: dark density}. 

\section{Bright Matter-Wave Solitons}

\subsection{Full perturbative approach}

In the case of attractive interactions ($s=-1$), Eq. (\ref{eq: GP perturbed}) 
reads  after substitution of $\tau=t/2$
\begin{equation}
i\partial_\tau u+\partial^2_x u+2|u|^2 u =2\epsilon P(u). \label{eq: bright GP perturbed}
\end{equation}
In the absence of perturbations, it is well known that Eq. (\ref{eq: bright GP perturbed}) 
possesses a bright soliton solution of the form
\begin{equation}
u(z,t)=2i \eta \exp(-2i\xi x-i\Phi) {\rm sech}(z), 
\label{eq: bright Ansatz}
\end{equation}
where $z=2\eta(x-\zeta)$, while $\eta$ represents the amplitude, 
$\Phi$ the phase and $\zeta$ the center of the soliton, and $\xi$ 
is related to 
the velocity of the soliton. For a small perturbation, we can now 
employ the adiabatic perturbation theory for bright solitons \cite{Kivshar89} 
to treat the perturbation effects analytically. Then, the soliton parameters 
become slowly-varying functions of time, however the shape of the soliton 
remains unchanged (once again this is a principal assumption that will
be justified a posteriori). With the general perturbation arising due to 
a spatially-dependent scattering length, one arrives at 
the following system of ordinary differential equations for the parameters of 
the soliton:
\begin{eqnarray}
 \partial_\tau \eta&=&8 \eta^2\xi\int dx\sqrt{g}\partial_x\frac{1}{\sqrt{g}}   
{\rm sech}^2(2\eta(x-\zeta))\label{eq: bright eta general},
\\
 \partial_\tau \xi &=&8\eta^3\int dx \sqrt{g}\partial_x\frac{1}{\sqrt{g}}  \tanh^2(2\eta(x-\zeta)) {\rm sech}^2(2\eta(x-\zeta))\nonumber\\
&&-2\eta^2\int dx \sqrt{g}  \partial_x^2 \frac{1}{\sqrt{g}} \tanh(2\eta(x-\zeta){\rm sech}^2(2\eta(x-\zeta)), 
\label{eq: bright xi general} \\
 \partial_\tau \zeta &=&-4\xi+8\eta\xi \int dx\sqrt{g}\partial_x\frac{1}{\sqrt{g}} (x-\zeta) {\rm sech}^2(2\eta(x-\zeta)), 
\label{eq: bright zeta general}\\
 \partial_\tau \Phi &=&4(\xi^2-\eta^2)+8\eta^2 \int dx \sqrt{g}\partial_x\frac{1}{\sqrt{g}} {\rm sech}^2(2\eta(x-\zeta))\tanh(2\eta(x-\zeta))
\nonumber\\
&&(1-2\eta x \tanh(2\eta(x-\zeta)))\nonumber\\
&&-2\eta \int dx \sqrt{g}  \partial_x^2 \frac{1}{\sqrt{g}} {\rm sech}^2(2\eta(x-\zeta))\bigr(1-2\eta x \tanh(2\eta(x-\zeta))\bigl). 
\label{eq: bright phi general}
\end{eqnarray}
For an interaction parameter of the form (\ref{g(x)}) one obtains:
\begin{eqnarray}
 \partial_\tau \eta&=&8\eta^2\xi\int dx \frac{C}{D+Cx}  {\rm sech}^2(2\eta(x-\zeta))\label{eq: bright eta}\\
 \partial_\tau \xi &=&8\eta^3\int dx \frac{C}{D+Cx}  \tanh^2(2\eta(x-\zeta)) 
{\rm sech}^2(2\eta(x-\zeta)) \label{eq: bright xi} \\
 \partial_\tau \zeta &=&-4\xi+8\eta\xi \int dx\frac{C}{D+Cx}(x-\zeta) 
{\rm sech}^2(2\eta(x-\zeta)) \label{eq: bright zeta}\\
 \partial_\tau \Phi &=&4(\xi^2-\eta^2)+8\eta^2 \int dx \frac{C}{D+Cx} 
{\rm sech}^2(2\eta(x-\zeta))\tanh(2\eta(x-\zeta))\nonumber\\
&&(1-2\eta x \tanh(2\eta(x-\zeta))). \label{eq: bright phi}
\end{eqnarray}

\subsection{Approximations}

From the above equations it is clear that Eq. (\ref{eq: bright phi}) describes the time evolution of the phase of the 
soliton which, however, 
does not emerge in the equations determining the other parameters. Therefore, we will 
restrict our considerations to Eqs. (\ref{eq: bright eta})-(\ref{eq: bright zeta}).
Since the soliton is exponentially localized around $x=\zeta$ we can 
perform a Taylor expansion around $\zeta$ and thus simplify 
Eqs. (\ref{eq: bright eta general})-(\ref{eq: bright zeta general}) as follows:
\begin{eqnarray}
 \partial_\tau \eta &=&8\eta\xi\sqrt{g(\zeta)}\partial_x \frac{1}{\sqrt{g(x)}}\Bigm|_{x=\zeta}
\label{eq: bright taylor eta general}\\
 \partial_\tau \xi&=&\frac{8}{3}\eta^2\sqrt{g(\zeta)}\partial_x \frac{1}{\sqrt{g(x)}}\Bigm|_{x=\zeta}
\label{eq: bright taylor xi general}\\
 \partial_\tau \zeta&=&-4\xi.
\label{eq: bright taylor zeta general}
\end{eqnarray}
The physical interpretation of this approximation is that the interaction 
parameter does not vary over the width of the soliton. 
Focusing more specifically on a collisional inhomogeneity of the 
form of Eq. (\ref{g(x)}) where these contributions vanish exactly, leads to
\begin{eqnarray}
 \partial_\tau \eta &=&\eta\xi\frac{8C}{D+C\zeta}
\label{eq: bright taylor eta}\\
 \partial_\tau \xi&=&\frac{8}{3}\eta^2\frac{C}{D+C\zeta} \label{eq: bright taylor xi}\\
 \partial_\tau \zeta&=&-4\xi.
\label{eq: bright taylor zeta}
\end{eqnarray}
We can solve the simplified Eq. (\ref{eq: bright taylor eta}) directly by 
using Eq. (\ref{eq: bright taylor zeta}):
\begin{eqnarray}
\eta&=&\eta(0)\frac{(C\zeta(0)+D)^2}{(C\zeta+D)^2}. 
\label{eq: bright zeta(tau)}
\end{eqnarray}
Combination of Eqs. (\ref{eq: bright taylor xi}-\ref{eq: bright zeta(tau)}) 
and back transformation to the real time $t$ leads to the equation of motion for the soliton center:
\begin{eqnarray}
 \partial^2_t \zeta=-\frac{8}{3}C\eta(0)^2\frac{(C\zeta(0)+D)^4}{(C\zeta+D)^5}, 
\label{eq: bright EOM}
\end{eqnarray}
%
%
with an associated effective potential:
\begin{eqnarray}
 V^{eff}(\zeta)=\frac{2}{3} \eta(0)^2 \frac{(C\zeta(0)+D)^4}{(C\zeta+D)^4}. \label{eq: birght effective pot}
\end{eqnarray}

\subsection{Numerical Results}

\begin{figure}[htbp]
\includegraphics[angle=270,width=7cm]{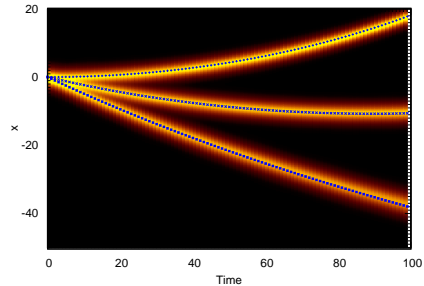}  
\caption{
Time evolution of the density of a bright soliton 
with initial parameter $\eta=0.5$, $\Phi=\zeta=0$ and $\xi_{init}=0$, $0.25$, 
$0.5$ (from top to bottom). The dotted line shows the result for $\zeta$ (the 
center of the soliton) from the adiabatic perturbation theory.} 
\label{fig: bright density}
\end{figure}

Figure \ref{fig: bright density} shows the time evolution of the density of the 
bright soliton (note that, in contrary to before, black represent zero density, while white represents high
density). The results are obtained by integration of Eq. 
(\ref{eq: GP perturbed}) with an initial state given by Eq. 
(\ref{eq: bright Ansatz}) with parameters initialized as
$\eta=0.5$, $\Phi=\zeta=0$ and $\xi_{init}=0$, $0.25$, $0.5$ 
(from top to bottom). The dotted line shows the corresponding results for 
the center of the soliton $\zeta$ obtained by solving 
Eqs. (\ref{eq: bright eta}-\ref{eq: bright phi}). The results of the 
perturbation theory once again agree very well with the results of the PDE. 
In the case of zero initial velocity (top curve), the soliton gets accelerated 
to the positive half-plane and starts moving into this direction immediately. 
For positive $\xi_{init}$ the initial velocity is negative, 
according to Eq. (\ref{eq: bright xi}), leading to a motion towards the 
negative half-plane. However, the soliton still 
moves toward the direction of the positive half-plane, due to its initial speed, 
yet eventually it acquires a zero velocity and a change of the direction of motion 
occurs. 
The direction of the acceleration is the direction of increasing interaction 
parameter as can be seen by comparing the results with Fig. \ref{fig: g(x)}, 
as this minimizes the energy of the system (even though the interaction does 
not act, strictly speaking, as a potential). 

\begin{figure}[htbp]
  \begin{minipage}[c]{7 cm}
\includegraphics[angle=270,width=7cm]{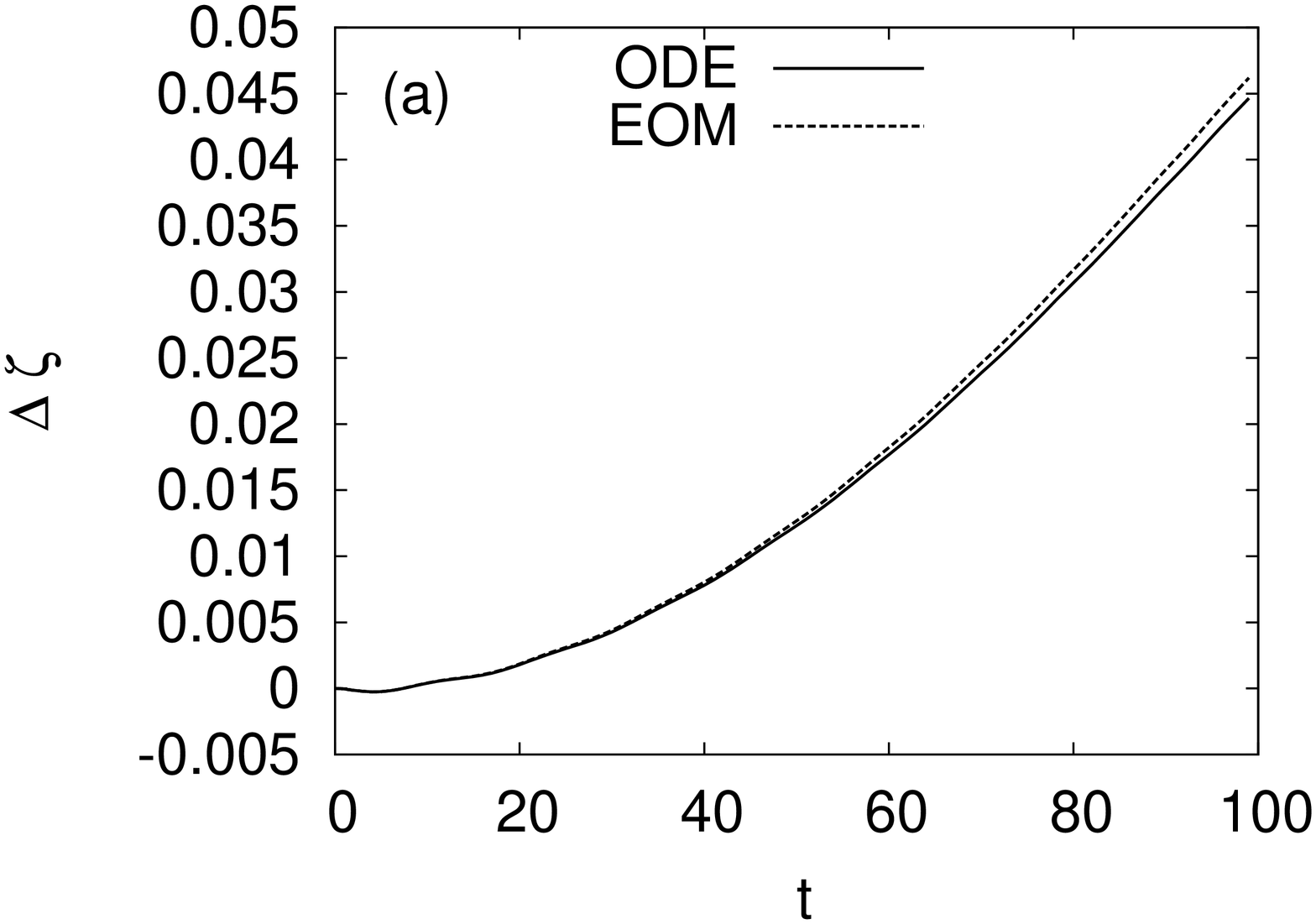}  
  \end{minipage}
  \begin{minipage}[c]{7 cm}
\includegraphics[angle=270,width=7cm]{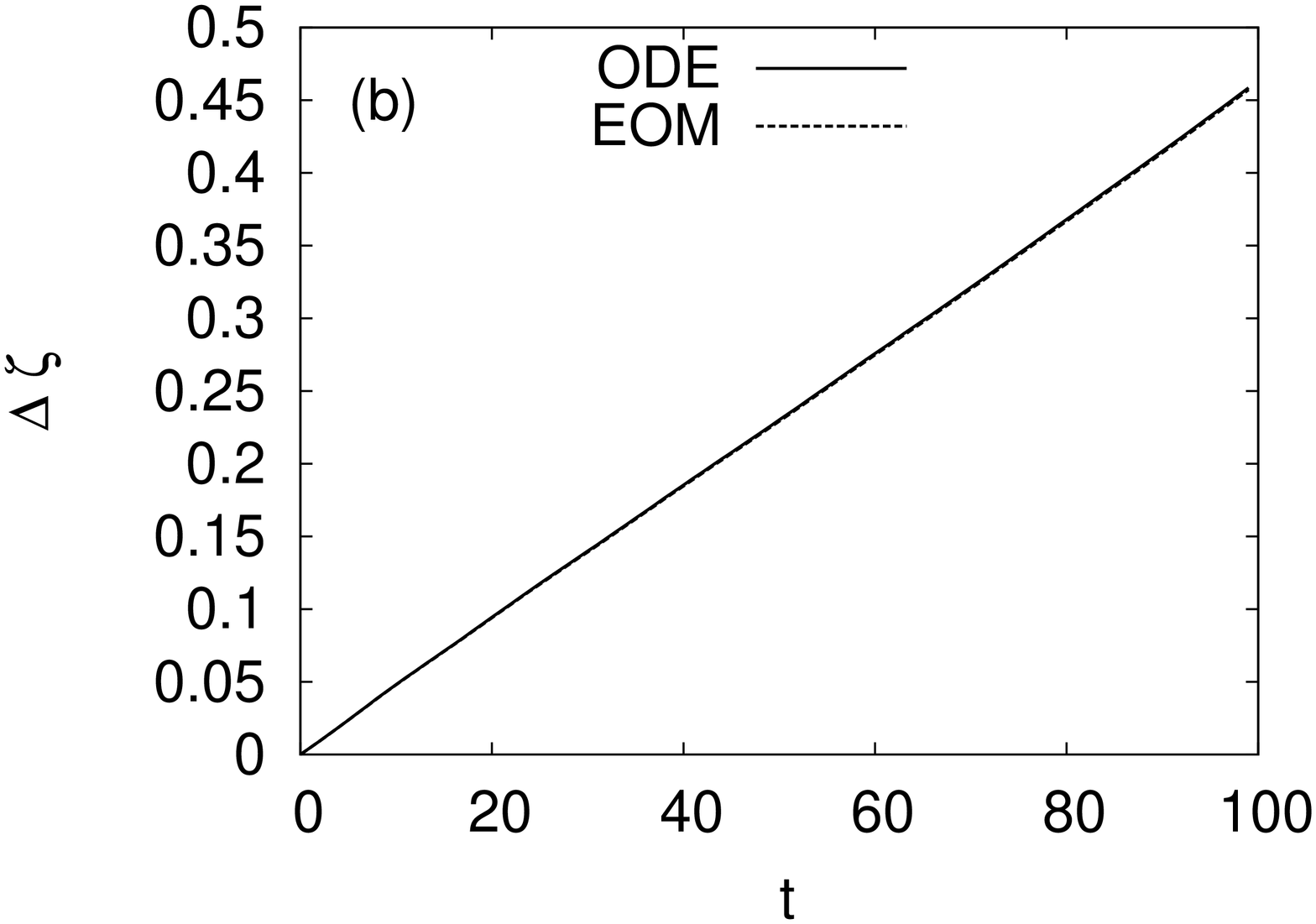}  
  \end{minipage}
  \caption{Difference of the soliton center calculated by solving the ODE and 
EOM to the result of the PDE with initial parameter $\eta=0.5$, $\Phi=\zeta=0$ 
and $\xi_{init}=0$ (a) or  $\xi_{init}=0.5$ (b). }
\label{fig: bright deltax}
\end{figure}

Figure \ref{fig: bright deltax} shows the difference of the center of the 
soliton calculated by the PDE with the results of the ODEs 
(\ref{eq: bright eta})-(\ref{eq: bright zeta}) and the EOM 
(\ref{eq: bright EOM}). The center of the soliton of the PDE solution is 
determined by the quotient $\zeta=\int x |u|^2 dx / \int |u|^2 dx $. 
Figure \ref{fig: bright deltax}a shows the differences for $\eta=0.5$, 
$\Phi=\zeta=0$ and $\xi_{init}=0$. The difference of the EOM result from
the PDE result is slightly larger than the difference of the ODE result. 
Both differences increase with time 
but they are still very small for 
the time period considered. Fig. \ref{fig: bright deltax}b shows the 
differences for $\eta=0.5$, $\Phi=\zeta=0$ and $\xi_{init}=0.5$. In this case,  
the absolute differences are an order of magnitude larger than in the previous 
case. However, compared to the position and width of the soliton one can still 
regard them as small. The results of the ODEs and the EOM are almost equal. 
A conclusion of this investigation is that the dynamics of the soliton is described 
fairly accurately by the model of a particle subject to the effective 
potential (\ref{eq: birght effective pot}).

\section{Conclusions and Future Challenges}

In this work, we considered the effect of (slowly-varying) spatially
dependent nonlinearities of a definite sign on both dark and bright matter-wave solitons 
of repulsive and attractive Bose-Einstein condensates, respectively. 
We have shown that a relevant transformation can be 
employed to convert the spatially dependent nonlinear problem into one of 
spatially uniform nonlinearity, at the expense of introducing 
two perturbative terms. One of the latter is in the form of a 
linear potential (which have been considered extensively previously),
while the other constitutes a non-potential type of perturbation,
being proportional to the spatial derivative of the field. To
especially highlight the non-potential nature of the second term,
we considered collisional inhomogeneities of inverse square spatial
dependence, whereby the linear potential perturbation identically vanishes,
and the purely non-potential one has to be considered. Even in these
settings (but also more generally), we found that soliton perturbation
theory provides a powerful tool towards describing such collisional
inhomogeneities.

It would be interesting to extend the present considerations in a 
number of directions. Firstly, it would be relevant to appreciate 
the effect of $g(x)$ on higher-dimensional structures, such as 
vortices, and on their stability. On the other hand, it would
be especially interesting even in one spatial dimension to determine
whether techniques like the ones used here (or variants thereof) can
be applied to cases where the sign of the nonlinearity changes.
Finally, it would be relevant to observe systematically how techniques such 
as soliton perturbation theory may fail, as the size of spatial extent
of the collisional inhomogeneity becomes comparable to that of the 
solitary wave and to understand the ensuing phenomenology in such cases.
Studies along some of these directions are currently in progress and
will be reported in future publications.

\vspace{5mm}

{\bf Acknowledgements}. PGK gratefully acknowledges
support from NSF-DMS-0349023, NSF-DMS-0505663, NSF-DMS-0619492,
NSF-DMS-0806762 and from the Alexander von Humboldt Foundation. 
The work of DJF was partially supported by the Special Research 
Account of the University of Athens. S.M. gratefully appreciates
financial support by the Heidelberg Graduate School of Fundamental
Physics in the framework of a visit to the university of Massachusetts at Amherst.


\end{document}